\documentclass[apj]{emulateapj}
\journalinfo{\textsc{The Astrophysical Journal}}
\slugcomment{Received 2015 April 04; Accepted 2015 June 24}
\usepackage{url}

\usepackage{apjfonts}
\usepackage{natbib}
\usepackage{booktabs}
\usepackage{ThreePartTable}
\usepackage{epstopdf}
\usepackage{graphicx}

\usepackage[colorlinks,
        hypertexnames=false,
        pagebackref=true,
        linkcolor=blue,    % color of internal links
    citecolor=blue,        % color of links to bibliography
    filecolor=magenta,     % color of file links
    urlcolor=blue          % color of url
]{hyperref}                %   package that links in pdf
\usepackage{amsmath}
\shorttitle{Consistency of twist in a magnetic cloud and its solar source}
\shortauthors{Vemareddy et al}

\begin{document}
\title{Comparison of magnetic properties in a magnetic cloud and its solar source on April 11-14 2013 }
\author{P.~Vemareddy$^1$, C.~M{\"o}stl$^{2,3}$, T.~Rollett$^2$, W.~Mishra$^4$, C.~Farrugia$^5$, and M.~Leitner$^3$}
\affil{$^1$Indian Institute of Astrophysics, Koramangala, Bangalore-560034, India}
\affil{$^2$Space Research Institute, Austrian Academy of Sciences, A-8042 Graz, Austria} 
\affil{$^3$IGAM-Kanzelh{\"o}he Observatory, Institute of Physics, University of Graz, A-8010 Graz, Austria}
\affil{$^4$Department of Geophysics and Planetary Sciences, University of Science and Technology of China, Hefei-230026, China}
\affil{$^5$Space Science Center and Department of Physics, University of New Hampshire, Durham, NH, USA}
\email{vemareddy@iiap.res.in}
%%%%%%%%%%%%%%%%%%%%%%%%%%%%%%%%%%%%%%%%%%%%%%%%%%%%%
%% Abstract %
%%%%%%%%%%%%%%%%%%%%%%%%%%%%%%%%%%%%%%%%%%%%%%%%%%%%%
\begin{abstract}
In the context of Sun-Earth connection of coronal mass ejections and magnetic flux ropes (MFRs), we studied the solar active region (AR) and the magnetic properties of magnetic cloud (MC) event during April 14-15, 2013. We use in-situ observations from the Advanced Composition Explorer and source AR measurements from the Solar Dynamic Observatory. The MCs magnetic structure is reconstructed from the Grad-Shafranov method which reveals a northern component of the axial field with left-handed helicity. The MC invariant axis is highly inclined to the ecliptic plane pointing northward and is rotated by $117^o$ with respect to the source region PIL. The net axial flux and current in the MC are comparatively higher than from the source region. Linear force-free alpha distribution ($10^{-7}-10^{-6}$ m$^{-1}$) at the sigmoid leg matches the range of twist number in the MC of 1-2 AU MFR. The MFR is non-linear force-free with decreasing twist from the axis (9 turns/AU) towards the edge. Therefore Gold-Hoyle (GH) configuration, assuming a constant twist, is more consistent with the MC structure than the Lundquist configuration of increasing twist from the axis to boundary. As an indication to that, the GH configuration yields better fitting to the global trend of in-situ magnetic field components, in terms of rms, than the Lundquist model. These cylindrical configurations improved the MC fitting results when considered the effect of self-similar expansion of MFR. For such twisting behaviour, this study suggests an alternative fitting procedure to better characterise the MC magnetic structure and its source region links.
\end{abstract}

\keywords{Sun:  heliosphere--- Sun: flares --- Sun: coronal mass ejection --- Sun: magnetic fields---
Sun: filament --- Sun: solar-terrestrial relations}
%%%%%%%%%%%%%%%%%%%%%%%%%%%%%%%%%%%%%%%%%%%%%%%%%%%%%%%%%
%% 1. Introduction %
%%%%%%%%%%%%%%%%%%%%%%%%%%%%%%%%%%%%%%%%%%%%%%%%%%%%%%%%%
\section{Introduction}
\label{Intro}
Magnetic clouds (MCs) are large scale, organized magnetic structures in interplanetary space \citep{burlaga1981}, characterized by a smoothly rotating field of enhanced field strength, low proton temperature, and low proton beta. They are usually observed in situ as interplanetary coronal mass ejection (ICMEs), that are generally preceded by the occurrence of major coronal mass ejections (CMEs) at the Sun. Many ICMEs are likely to be associated with an MC depending on the trajectory of spacecraft \citep{xieh2013}. It is now believed and shown from a variety of independent studies that MCs are magnetic flux ropes (MFRs) of locally straight cylindrical geometry \citep{burlaga1988, farrugia1995, shodhan2000, liuying2008, gopalswamy2013, huqiang2014}. In this picture, the MC is thought to be part of a large-scale bent flux rope extending from the Sun into interplanetary space with its feet possibly still connected to the Sun \citep{burlaga1991,farrugia1993, bothmer1998, webb2000}. From this point of Sun-Earth connection, a major interesting, important question is how the solar source regions are connected to the in situ MCs, which should lead to important clues on how to forecast the internal magnetic field of CMEs around Earth and other planets.

Towards this scientific aspect, MCs are mostly studied from in situ one-dimensional observations.Various fitting models have been developed to reconstruct the global picture of MCs in two or three dimensions (e.g., \citet{lepping1990, huqiang2002,al_haddad2013,janvier2015}). On the other hand, the solar source regions are studied for the onset of the CME \citep{moore2001, forbes2006, kliem2006, chenpf2011, chengx2012, vemareddy2012a, vemareddy2014b} and its propagation is tracked from Sun to near Earth using various space based observations to confirm the connection of solar source regions with MCs \citep{gopalswamy2001, manoharan2006, davies2009,liuying2010a,liuying2010b,temmer2012, harrison2012, mostl2012, liuying2013,webb2013,mostl2014,vemareddy2015b}. Although, at the two ends, we thus enhanced our understanding in connecting MCs at 1AU to the Sun in many cases \citep{gopalswamy2013, xieh2013}, relating the source region signatures to magnetic properties in MCs remain still lacking.

The commonly recognized plasma structures as flux ropes in magnetically active regions (ARs) on the Sun include filaments, sigmoids and erupting loops. As direct magnetic observations of the coronal flux ropes are not possible, the amount and distribution of twist in the flux ropes are inferred generally, as a proxy, from photospheric magnetic field observations \citep{pevtsov1995, hagino2004, vemareddy2012b}. Furthermore, there is a great difficulty in identifying an unambiguous one-to-one association between the MC and its solar progenitor due to CME-CME interactions \citep{gopalswamy2001b,burlaga2002,lugaz2005,mostl2012, liuying2012, liuying2014a, liuying2014c, harrison2012,  temmer2012,mishra2015}, CME-deflection \citep{wangyuming2004, wangyuming2014,liewer2015,mostl2015}, or multiple Earth facing ARs. Therefore, before associating the magnetic signatures from the solar source region to the in situ MC, it has to be assured that the MC is uniquely identified from its source region.      

Given the model of flux rope configuration to the in situ MCs \citep{lepping1990}, the total field line lengths at the MC boundary must be larger than at the center. This is a fundamental physical point to help assess the flux rope model of Sun-Earth connecting MCs. Based on this point, following the approach of \citet{larson1997}, \citet{kahler2011a,kahler2011b} compared the total field line lengths derived from the energetic electron beam spectrum, with that derived from Lundquist and flux conservation models. Their comparison, in a set of WIND MC events, implied a poor correlation between measured and modelled field line lengths, indicating doubt on the Lundquist flux rope concept to the MCs. Recently, this issue has been further clarified by the study of \citet{huqiang2015}, where they employed the Grad-Shafranov (GS;  \citealt{huqiang2002}) reconstruction technique for the MC's magnetic structure. It had been shown that the MC's magnetic structure is more aligned with constant twist Gold-Hoyle model \citep{goldhoyle1960} but not with Lundquist model which features increasing twist from the axis of the flux rope to its boundary. In addition, there were difficulties to find comparable magnetic parameters in the MC magnetic structure and its source regions. In a study of twelve interplanetary MCs, \citet{leamon2004} found the MC magnetic flux to be comparable to that of the associated AR. They used cylindrically symmetric constant $\alpha$ Lundquist model to derive the field line twist, total current, total magnetic flux from in situ observations of MC. However, the total field twists of the MCs were about an order of magnitude larger than those of the ARs. These findings led them to believe that MCs associated with AR eruptions are formed by magnetic reconnection between these regions and their larger-scale surroundings, rather than simple eruption of preexisting structures in the corona or chromosphere.

In the present paper, we study the source region magnetic properties of an MC event observed on April 14, 2013 using in situ observations from the Advanced Composition Explorer (ACE) and solar source AR measurements from the Solar Dynamic Observatory (SDO). Studies of its solar origins on initiation, eruption mechanisms unambiguously revealed build-up and onset of a sigmoidal flux rope to a large scale CME eruption facing Earth from AR 11719 \citep{vemareddy2014b}. Further connections of this CME eruption to our MC of interest are uncovered by another detailed study by tracking the CME to near Earth environment \citep{vemareddy2015b}. 

Motivated by many previous studies (e.g., \citealt{larson1997, leamon2004, mostl2009, kahler2011a, kahler2011b, huqiang2014, huqiang2015}), in an attempt to provide further details about the in situ magnetic structure of flux ropes and their Sun-Earth connections of major solar eruptions, we employ the GS reconstruction technique and then compare the MC's orientation and magnetic signatures with its solar source region \citep{liuying2010b,nieveschinchilla2012,liuying2016}. Moreover, the GS reconstruction results are compared with the cylindrically symmetric linear force-free Lundquist model and the non-linear force-free Gold-Hoyle model fitting results in order to assess the differences in the estimated fitting parameters. In Section~\ref{sec2}, we present the observations of the solar source region of the magnetic flux rope. The results of GS reconstruction from in situ MC observations are described in Section~\ref{sec3}, and its magnetic signatures are compared to that from the source region in Section~\ref{sec4}. In section~\ref{sec5}, the MC fitting based models are discussed in conjunction with GS reconstruction. Conclusions are highlighted with a possible discussion in Section~\ref{sec6}.   
\begin{figure*}[!ht]
\centering
\includegraphics[width=.99\textwidth,clip=]{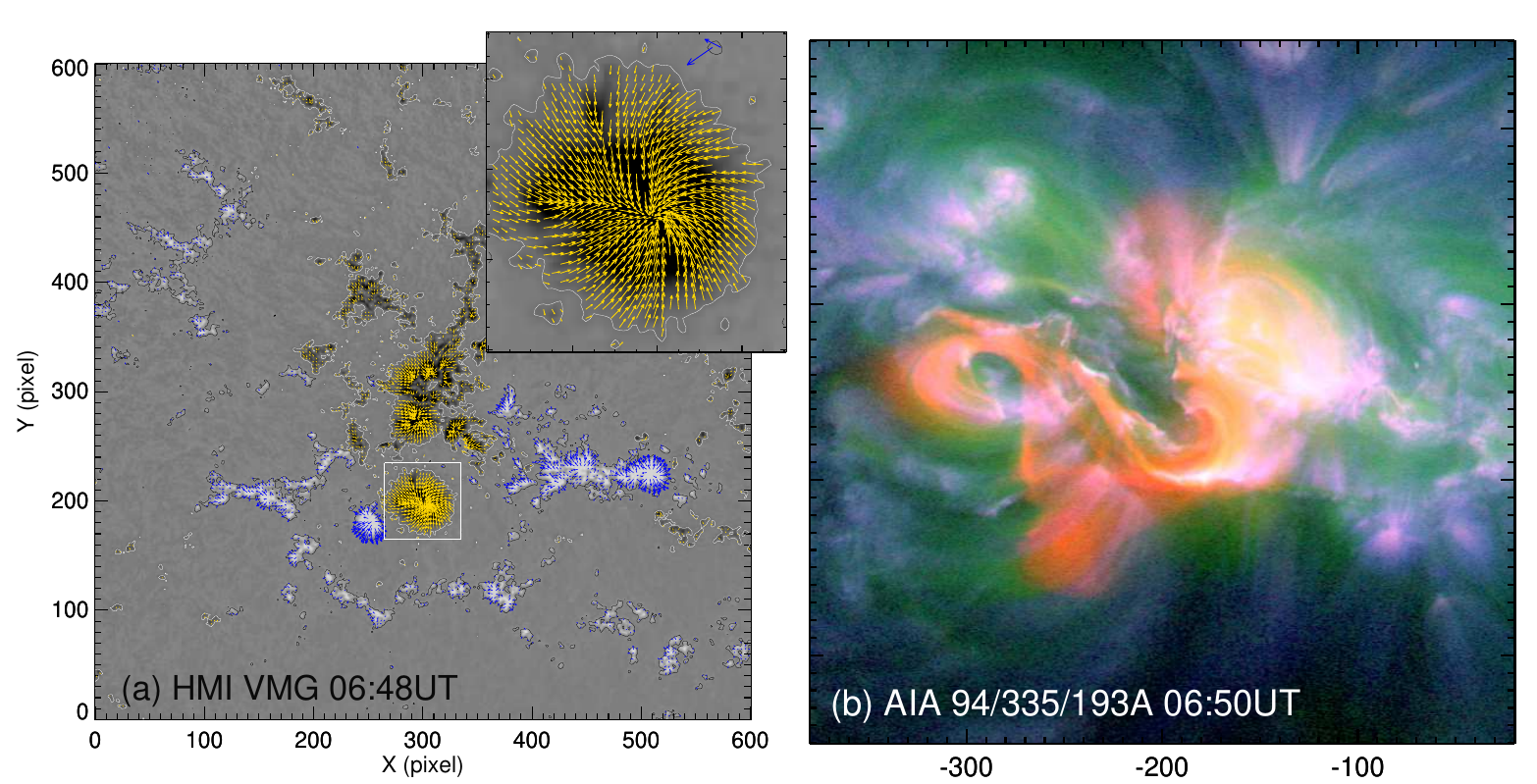}
\caption{Observations of source AR 11719 on the Sun on April 11, 2013. (a) HMI Vector magnetic field measurements plotted as transverse vectors (arrows) on the vertical magnetic field map. Inset plot is a blow up of vector magnetic field in the negative polarity sunspot from which the sigmoid originates. Note that orientation of the transverse vectors is in left handed direction. (b) Composite image of the sigmoid (MFR) prepared from simultaneous AIA 94 (red), 335 (green), 193 (blue)~\AA~channel observations. Axis coordinates in the right panel are in heliocentric arcsec units. }
\label{Fig1}
\end{figure*}

\section{Source Active Region of CME Eruption}
\label{sec2}
The MC's source region on the Sun is AR 11719. From this AR, a halo CME eruption occurred on April 11, 2013 at 06:50UT. At this moment, the AR is located at N9E13. The AR consists of a filament channel which is overlaid by an inverse-S sigmoid (see Figure~\ref{Fig1}). Regarding this sigmoid as an MFR system, exceedingly critical twist in the MFR (kink instability) is interpreted as an initiation mechanism of the eruption at 06:50UT preceded by a GOES class M6.6 flare (see more details in \citealt{vemareddy2014b}). Moreover, a torus instability \citep{torok2005} is evidenced as a later mechanism to further drive the CME eruption. 

Figure~\ref{Fig1}(left panel) shows the vector magnetic field measurements of the AR by Helioseismic Magnetic Imager (HMI; \citealt{schou2012, hoeksema2014,centeno2014}) on board SDO. The AR consists of a main sunspot (inset)  from which the sigmoid (right panel) originates and lies along the polarity inversion line between negative in the north and positive flux in the south. Magnetic fields are evolving with decreasing flux content in both polarities for three days preceding the eruption. Magnetic fields in the sunspot from which the sigmoid originates clearly show left-handed orientation. Using these vector magnetic field observations, we calculate the average value of the force free parameter ($\alpha_{av}$) given by 
\begin{equation}
{{\alpha }_{av}}=\frac{\sum{{{J}_{z}}(x,y) sign[{{B}_{z}}(x,y)]}}{\sum{|{{B}_{z}}|}}
\end{equation}

where $J_z(x,y)=\frac{1}{{{\mu }_{0}}}{{\left( \nabla \times \mathbf{B}(x,y) \right)}_{z}}$ is the vertical current distribution. As the local distribution of $\alpha(x,y) (J_z(x,y)/B_z(x,y)) $ measures the extent of twist of the field lines due to field-aligned currents, its average $\alpha_{av}$ is a proxy to quantify the overall twist of entire AR magnetic structure \citep{hagino2004}. Its value at the start (00:00UT) of April 11, is calculated as $-0.3\times10^{-8}m^{-1}$  and found to increase predominantly to around $-2\times10^{-8}m^{-1}$ by 12:00UT on April 11.  The negative value of $\alpha_{av}$ indicates left-handed magnetic field twist in the AR and hence a dominant negative helicity. This is well consistent with the observed geometry of the inverse-S shaped coronal sigmoid.

\begin{figure*}
\centering
\includegraphics[width=.7\textwidth,clip=]{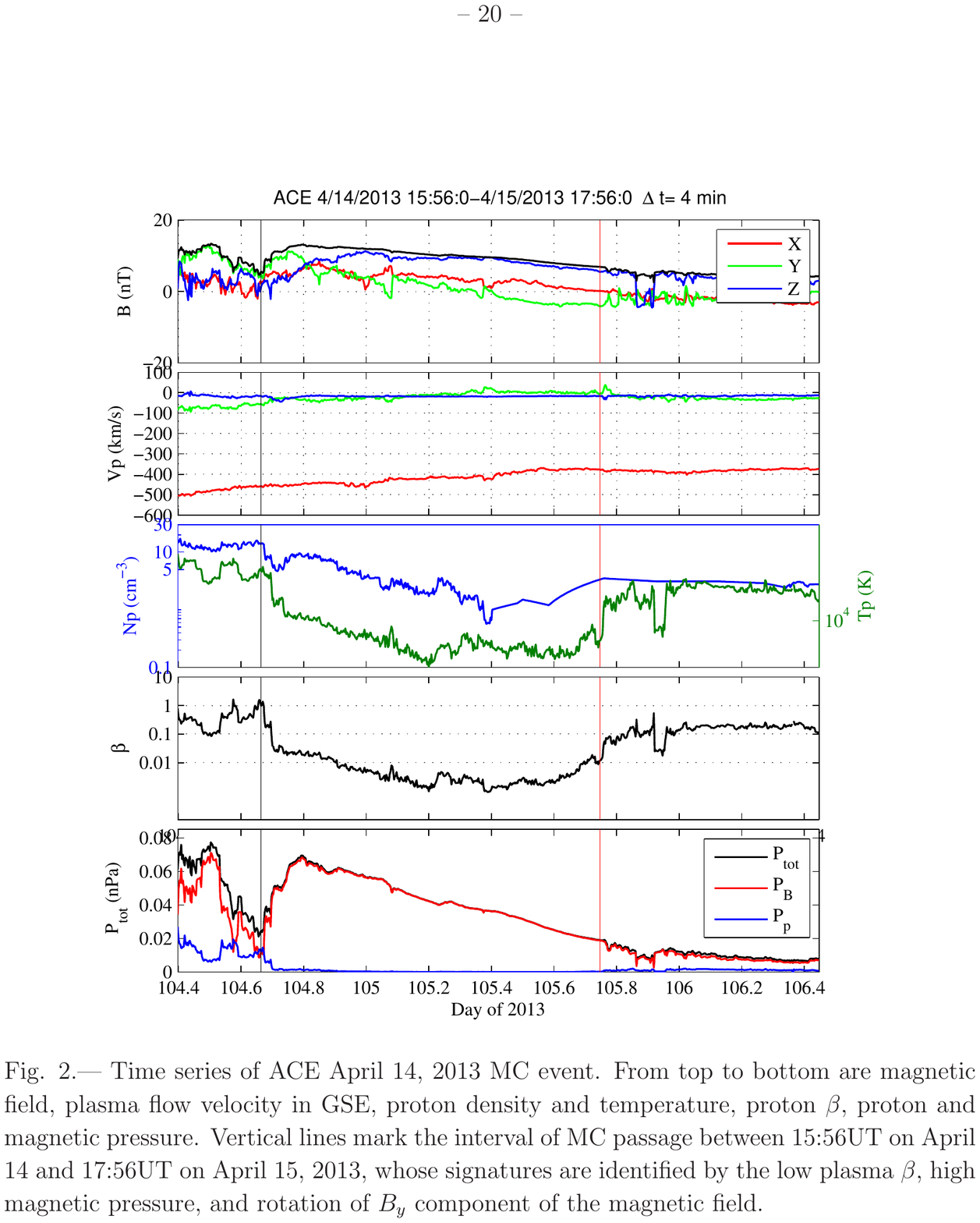}
\caption{Time series of ACE April 14, 2013 MC event. From top to bottom are magnetic field, plasma flow velocity in GSE, proton density and temperature, proton $\beta$, proton and magnetic pressure. Vertical lines mark the interval of MC passage between 15:56UT on April 14 and 17:56UT on April 15, 2013, whose signatures are identified by the low plasma $\beta$, high magnetic pressure, and rotation of $B_y$ component of the magnetic field. }
\label{Fig2}
\end{figure*}

As further studied in \citet{vemareddy2015b}, the disappearing net flux was suggested to help in sustaining and developing the sigmoid with increasing twist. Localized twist measurements from vector magnetograms also support the availability of critical twist (more than one turn within the arc length of the sigmoid, see Figure 7 in \citealt{vemareddy2014b}) in the sigmoid before the onset of the eruption. This critical twist is crucial for the onset of initial rise motion as a mechanism of the kink-instability \citep{torok2005}.

\begin{figure}[!htp]
\centering
\includegraphics[width=.49\textwidth,clip=]{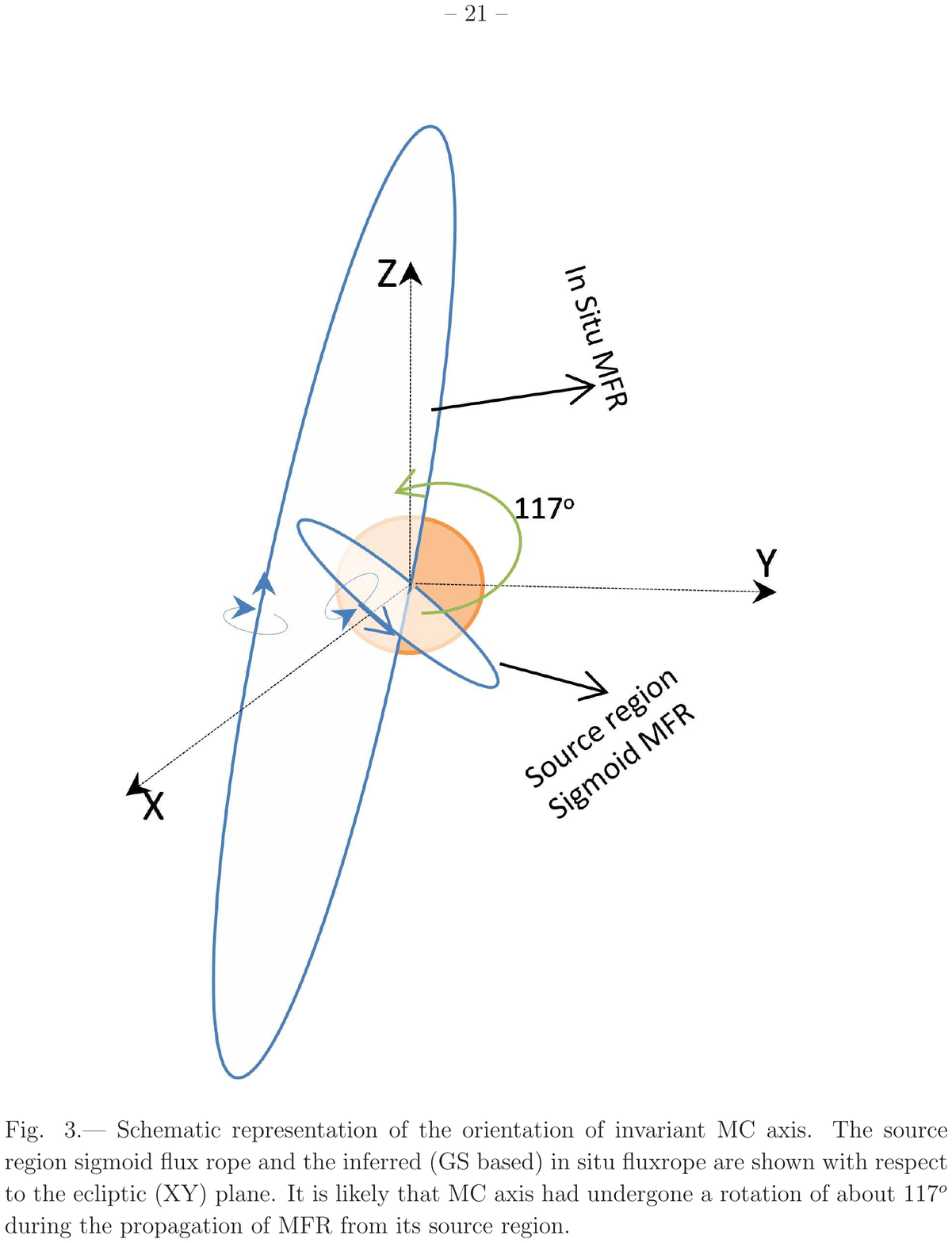}
\caption{Schematic representation of the orientation of invariant MC axis. The source region sigmoid flux rope and the inferred (GS based) in situ fluxrope are shown with respect to the ecliptic (XY) plane. It is likely that MC axis had undergone a rotation of about $117^o$ during the propagation of MFR from its source region. }
\label{schema}
\end{figure}

The tilt of the sigmoid (referred as MFR) is $45^o$ to the central meridian, i.e. the fluxrope axis makes an angle of roughly $225^o$ to the solar north in counter-clockwise direction (See schematic in Figure~\ref{schema}). This value of the tilt angle, in combination with other source region parameters, visually fits the observed CME morphology captured in the STEREO and LASCO field of view to a good extent \citep{vemareddy2015b}. As this CME is a halo and Earth directed, its arrival at L1 point is identified with a shock (on 13 April, 22:50UT), leading edge (on 14 April 14:35UT), and trailing edge (on 15 April 17:50UT) with the characteristics of a MC from the in situ velocity and magnetic field measurements. As studied in \citet{vemareddy2015b}, the in-situ parameters interpret a tilt angle of $360^o$ with respect to solar north in anti-clock direction. This mismatch in the orientation of the MFR (difference of roughly $135^o$ tilt angle) in the source region and in the in-situ MC could well be due to the rotation of the MFR apex  during its initiation and/or propagation \citep{liuying2010b,vourlidas2011}. As predicted by numerical and observational studies \citep{fany2003,green2007,lynch2009}, the apices of MFRs would rotate due to inherent twist in them, left handed helical ones in counter clockwise direction and right handed in clockwise direction. In our case of left handed MFR, a counterclockwise rotation is expected.  

As seen in Figure~\ref{Fig1}, the MFR axial field is pointing in the lower right (southwest), because MFR poloidal field is coming out of the photosphere in the south and going into the photosphere in the north of the magnetic neutral line. For such a poloidal field, a left handed MFR has an axis to the southwest. Therefore, the presumed MFR rotation would be counterclockwise roughly on the order of $135^o$ to match the in situ flux rope orientation (see Figure~\ref{schema}). Resolving the source region signatures of MFRs is essential for their Sun-Earth connections as observationally tracking the magnetic structure of the CME has not yet been made possible. 

\begin{figure*}[!htpb]
\centering
\includegraphics[width=.95\textwidth,clip=]{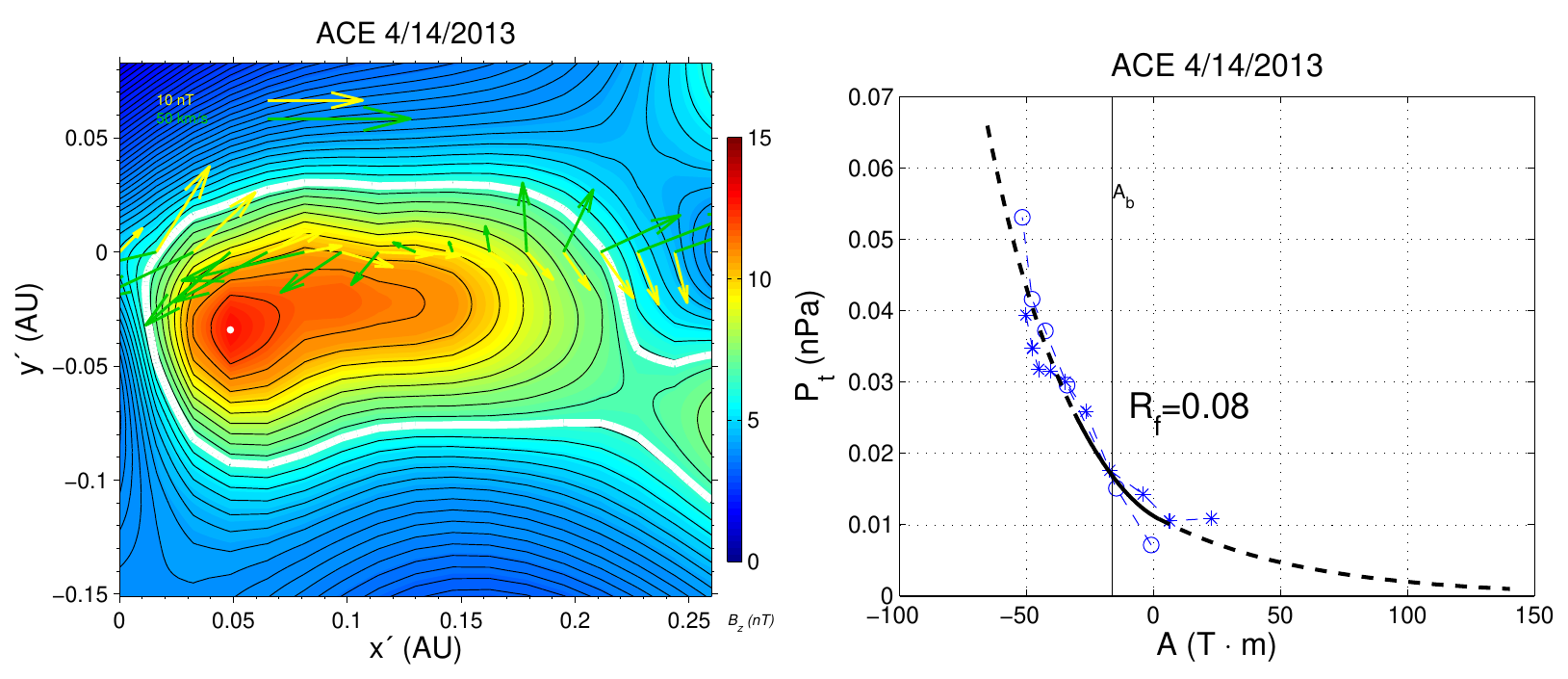}
\caption{{\bf Left:} Reconstruction of ACE magnetic cloud event. The cross section of the magnetic cloud is plotted as contours of magnetic vector potential (A) and filled contour plot of $B_z$ in color with scale shown by the colorbar. Yellow (green) arrows along $y=0$ line show transverse magnetic (velocity) field, indicating left handed chirality of the magnetic field in the flux rope associated with this magnetic cloud. White thick contour refers to the MC boundary used for calculating magnetic fluxes. {\bf Right:} The data plot and fitting curve for $P_t(A)$ of the magnetic cloud event. Scattered points are spacecraft measurements, circles for the first half of the event and stars for the second half. Solid curve represents the fitting function, $P_t(A)$.}
\label{Fig3}
\end{figure*}

\section{GS Reconstruction of MC Magnetic Structure}
\label{sec3}
We employed the GS reconstruction technique \citep{huqiang2002} to construct the magnetic field structure in the MC cross-section. This technique involves an assumption of translational symmetry of magnetic field along the fluxrope and so enables us to construct the field in a 2D cross-section with an invariant z-axis. The required in situ magnetic and velocity field observations (in GSE coordinate system) of the MC are obtained from ACE. In Figure~\ref{Fig2}, we show various observed parameters with time, during the MC passage through the spacecraft. A MC passage is generally identified by a strong magnetic field strength, low plasma $\beta$ (=$2{{\mu }_{0}}p/{{B}^{2}}$), rotation of any magnetic field component (reversal of sign during passage). As an outcome of the application of the GS method, we set the time interval between 15:56UT on April 14 [day of year as 104] and 17:56UT April 15 [day of year as 105], indicating its large size. Note the $B_y$ component changes sign from positive to negative, while $B_x$ and $B_z$ components remain positive during this MC passage. This means that this MC belongs to east-north-west (ENW) category according to the classification schemes introduced by \citet{bothmer1998} and \citet{mulligan1998}.

Recovering the magnetic field in the MC cross-section essentially involves determining the deHoffmannTeller (HT) frame and orientation of the MC axis \citep{huqiang2002}. Minimization of the mean square convective electric field in a moving frame gives the velocity of the HT frame ($V_{HT}$) as 414 km/s. Minimum variance analysis (MVA) of magnetic field vectors and construction of residue maps determined the exact MC axis orientation pointing at $72^o$ latitude ($\theta$) and $101^o$ longitude($\phi$). $\phi$ is the longitude, being $0^o$ towards GSE X, $+90^o$ towards GSE Y and so on. $\theta$ is the latitude, can also be called inclination, that is $0^o$ in the ecliptic, $+90^o$ towards ecliptic north (along GSE $+Z$) and $-90^o$ to ecliptic south (GSE $-Z$). We point that the latitude of $72^o$ refers to an alignment of MC axis with respect to the ecliptic. In other words, it is away from the ecliptic North (vertical plane to ecliptic) by $18^o$. This is consistent with the interpreting arguments by \citet{vemareddy2015b}, where they speculate a possible rotation of MFR apex up to $135^o$ due to inherent nature of handedness of magnetic field in the MFR. In this case, left handed helicity of magnetic field in the MFR might lead to a rotation in counterclockwise direction (as seen in line-of-sight) during its outward propagation, tending to align the MFR axis plane roughly perpendicular to the ecliptic plane. Particularly from the source region observations, the apex is likely rotated by about $117^o(45^o+72^o)$ to match the GS result of MC axis orientation. Schematic in Figure~\ref{schema} depicts the above described orientation of the in situ and source region MFR with respect to the ecliptic plane. We believe that the MC axis orientation is a crucial physical parameter and provides constraining clues on the Sun-Earth connection of MFRs.

Next, the measurements of transverse pressure $P_t(A)$, which is a sum of plasma pressure p and axial magnetic pressure $B_z^2/2\mu_0$, are re-sampled onto a grid of 17 points using an anti-aliasing re-sample function, and are then fitted with a second order polynomial as shown in Figure~\ref{Fig3}(right). A residue value $R_f= 0.08$ quantitatively describes the goodness of fit to the most data points. The resulting reconstructed map of MC's cross-section is plotted as contours of flux function (A) and filled contour plot of $B_z$ in color with a scale (left panel in Figure~\ref{Fig3}). The white thick contour refers to the MC boundary used for calculating the magnetic fluxes, which is based on the point of divergence of the inbound and outbound $P_t(A)$ functions at the value of $A_b$ in the right panel. This contains a slightly smaller interval than the MC interval as given by the boundaries in Figure~\ref{Fig2}. Closed contours of A represent helical field lines winding the axis in projection. The MC is larger than earlier studied cases \citep{huqiang2002, mostl2009} having a size of 0.26 AU. The closest distance of the spacecraft from the MC axis is 0.034AU on the positive y-side. This cross sectional map could be influenced by a possible MC expansion. Leading and trailing edges of MC are moving at a velocity of 461.4 and 376.7 km/s respectively. The ratio of MC expansion velocity $V_{exp}$($=$(leading-trailing) edge velocity/2) and $V_{HT}$ is 0.102, which is feeble to a significant expansion effect. 
The $B_z$ distribution is positive with a maximum field strength of 13 nT. The transverse ($B_x$, $B_y$ components) magnetic components show a left hand winding of field lines in the MC and therefore the helicity is negative consistent with the source region (Figure~\ref{Fig1}). GS result of various parameters of the MC are summarized in Table~\ref{tab1}. 

\begin{figure*}[!htpb]
\centering
\includegraphics[width=.95\textwidth,clip=]{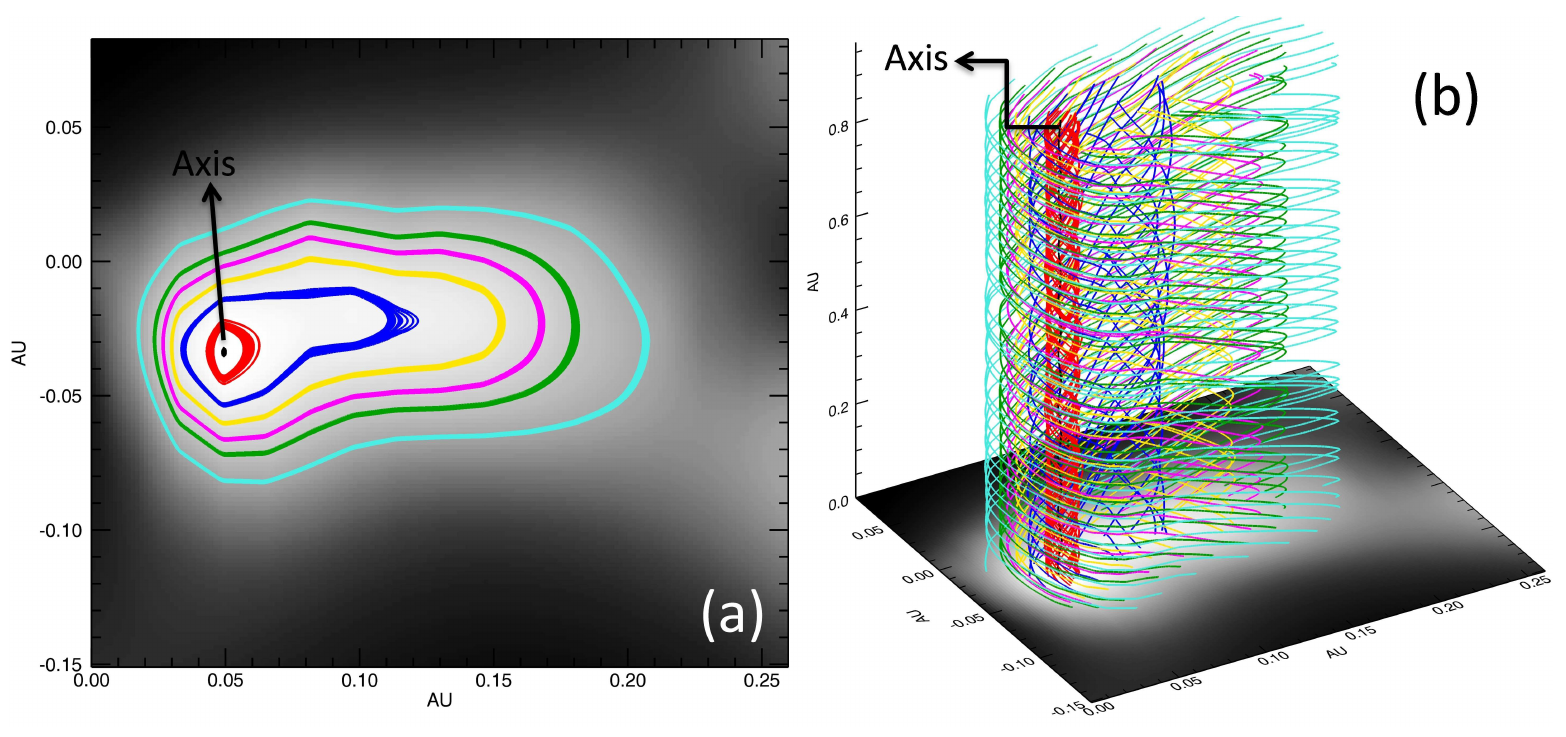}
%% command to convert jpg to eps files
%% bmeps -p=3 -c MC_FR.jpg MC_FR.eps 
\caption{Field lines in different flux shells of MC cross-section plotted in (a) top view, (b) perspective 3D view. Bottom boundary is $B_z$ map. Central field line is almost straight and represents the axis of the MFR. Vertical height is set such that every field line completes at least one full turn in order to enable winding number calculation. }
\label{Fig4}
\end{figure*}

Having three components of magnetic field in the MC cross-section, we can quantify the twist of the field lines \citep{huqiang2014}. As the closed contours of the flux function A represent field lines of one/more turns in the plane projection, the same can be realized by integrating field lines at any radial position from the axis of the MFR. The axial length (L) of any such field line which completes one full turn has a relation to the twist number
\begin{equation}
\tau =\frac{1}{L}turns/AU
\end{equation}

\begin{figure}[!htp]
\centering
\includegraphics[width=.49\textwidth,clip=]{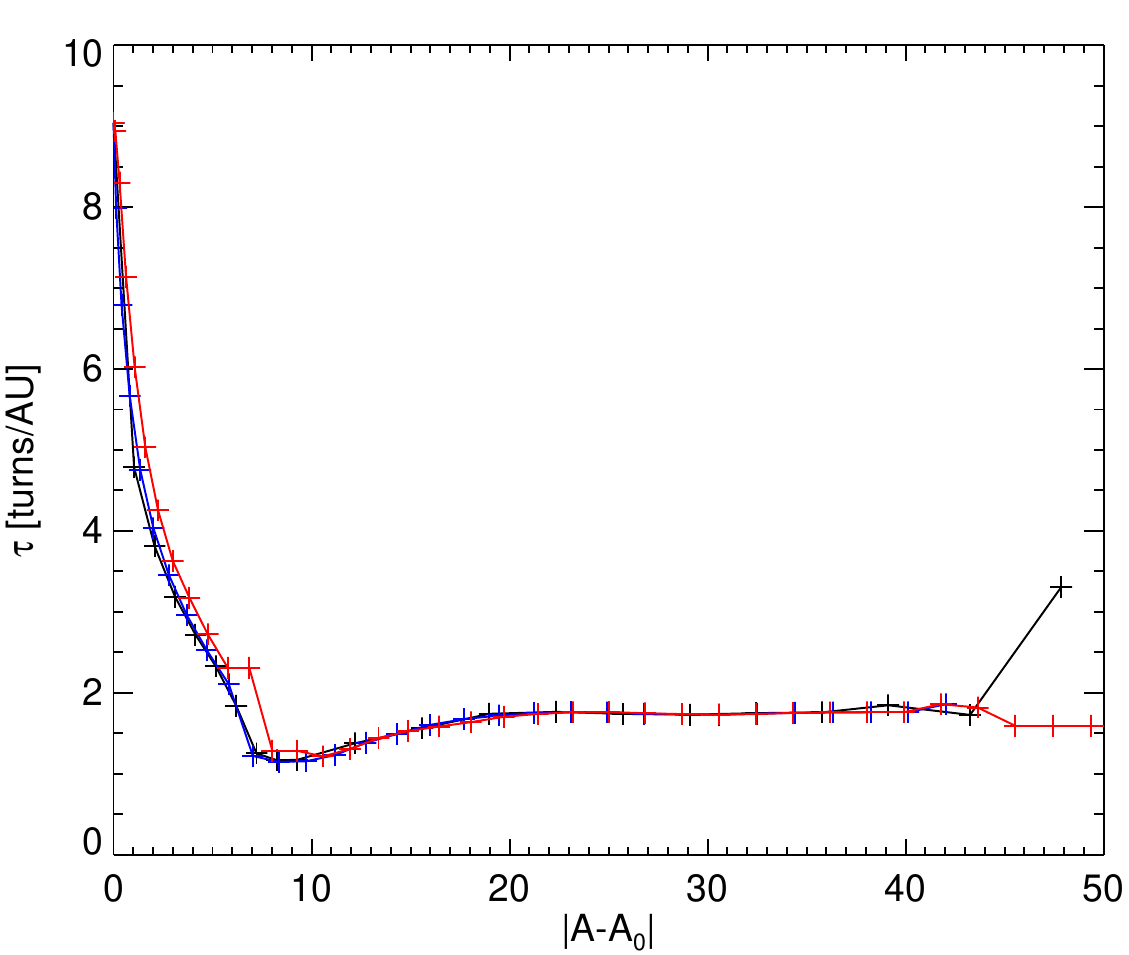}
\caption{Twist number as a function of shifted flux function in three different (each color for each direction) radial directions from the axis of the MC. Twist number rapidly falls from 9 turns/AU at the axis to stabilize to a constant twist number of 1.75 turns/AU upto the MC boundary.}
\label{Fig5}
\end{figure}

An involved assumption in the above equation is that the MC structure does not change, i.e., translational symmetry, along the full length of the rope. In Figure~\ref{Fig4}, we have plotted field lines in different radial positions from the center of the MC. By the nature of twist, the field lines wind about the axis in each flux shell. These field lines give an impression of a locally straight section of a large scale bent flux rope, which is constituted by flux shells. Note that the central field line (axis) is almost straight. We then measure the axial length of each field line that completes one full turn about the axis. The twist number in three different radial directions from the axis (i.e. $A_0$) is plotted as a function of shifted flux function $|A-A_0|$ in Figure~\ref{Fig5}. In the three different radial directions, starting from 9 turns/AU, $\tau$ declines rapidly from the center to 1.2 turns/AU in within a few units of flux function (radius). From there, the twist number shows slight increase to 1.75 turns/AU and continues further till the MC boundary. 

In this section, we present MC fitting results from two different cylindrical configuration in order to compare with GS results. 
\subsection{Gold-Hoyle versus Lundquist models}
We fitted the in situ magnetic field observations with the cylindrically symmetric linear force-free flux rope model \citep{lundquist1950} with radial functions of axial and poloidal field components

\begin{align}
  & {{B}_{z}}(r)={{B}_{0}}\text{Bessel}J(0,\alpha r) \nonumber \\
 & {{B}_{\varphi }}(r)=H{{B}_{0}}\text{Bessel}J(1,\alpha r)
\label{eq2}
\end{align}
where $\alpha$ is force-free parameter, H is the sign of helicity and $B_0$ is field strength. In this model, the magnetic field profile along the spacecraft's observational path is determined by the orientation of the flux rope axis i.e., z-axis which is given by the elevation and azimuth angle ($\theta$ and $\phi$), closest approach of the observational path p, flux rope diameter D, helicity sign H and field strength $B_0$ in the center of the flux rope \citep{lepping1990, leitner2007}. As a second flux-rope fitting model, we used the non-linear force-free Gold-Hoyle model (GH; \citealt{goldhoyle1960}) assuming a uniform field line twist (e.g. \citealt{farrugia1999, dasso2006, huqiang2015}). In this model, the radial functions of axial and poloidal field components are 
\begin{align}
  & {{B}_{z}}(r)=\frac{{{B}_{0}}}{1+{{T}_{0}}^{2}{{r}^{2}}} \nonumber \\
 & {{B}_{\varphi }}(r)=\frac{H{{T}_{0}}{{B}_{0}}r}{1+{{T}_{0}}^{2}{{r}^{2}}} 
\label{eq3}
\end{align}
where $T_0$ is twist number at the center. The fitting converges when $T_0=1.2$ turns/AU rather the GS result of  $T_0=9$ turns/AU because the rest of GH is totally consistent with GS.

\begin{figure*}[!htp]
\centering
\includegraphics[width=.7\textwidth,clip=]{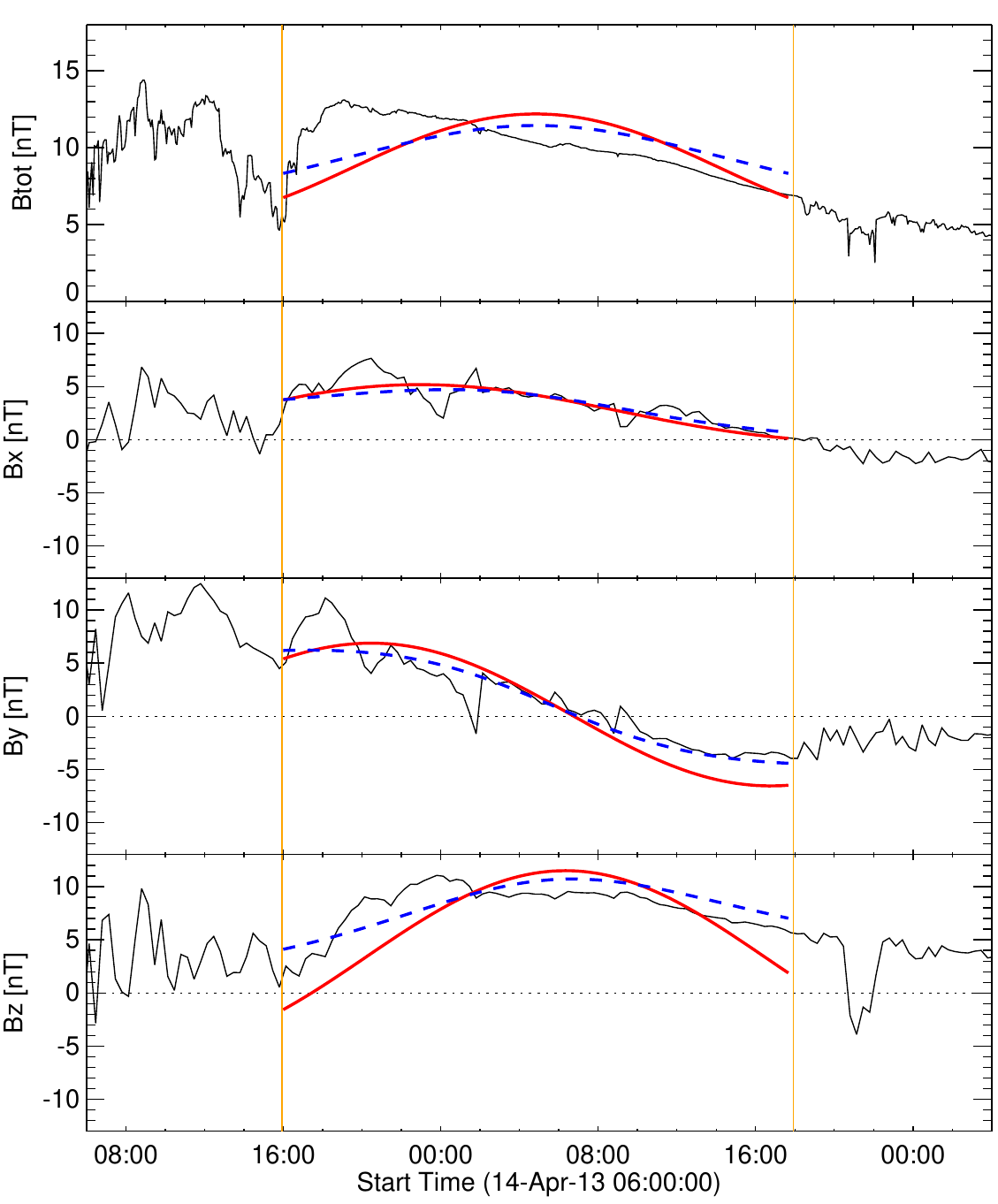}
\caption{Cylindrically symmetric constant $\alpha$ linear force-free Lundquist model fitting (solid red) and non-linear force-free GH (dashed blue) model fitting  are plotted against the in situ observations (black) in each panel. $T_0=1.2$ turns/AU is used for the GH configuration.The orange vertical lines mark the MC start and end times. Notice the better fitted rotating component with GH model.}
\label{Fig6}
\end{figure*}

Figure~\ref{Fig6} shows the results of the Lundquist model fitting (solid red) and the Gold-Hoyle model fitting (dashed blue) to the in situ magnetic field observations (black). The vertical lines mark the edges of the magnetic flux rope. As a measure of goodness of the fit, we compute the root mean square deviation (rms=$\sqrt{\sum\limits_{i=1,N}{{{\left( {{\mathbf{B}}_{o}}({{t}_{i}})-{{\mathbf{B}}_{m}}({{t}_{i}}) \right)}^{2}}}}/N$ ) between the observed magnetic field ($\mathbf{B}_o(t_i)$) and the modelled field ($\mathbf {B}_m(t_i)$) \citep{marubashi2007}. As can be noticed, both models reproduce the global trend of the field components, especially the rotating component, quite well. However, the GH configuration performs slightly better in terms of Erms parameter ($Erms=rms/max(\left|\mathbf{B}\right|)$). It is 0.266 for the Lundquist and 0.175 for the GH fit. The resulting orientation of the flux rope for both models is almost same. The helicity sign H and the field strength $B_0$, inclination angle are in agreement with the GS reconstruction results however, other parameters differ considerably. The derived parameters are listed against the GS values in Table~\ref{tab1}.

The radial profile of GS inferred $\tau$ is compared with that by the GH and Lundquist models (Figure~\ref{Fig5}). For this purpose, the GS derived field strength ($B_0=13nT$), twist ($T_0=9$ turns/AU=56.5radians/AU) and radius ($R_0=0.26 AU$) are supplied to Equations~\ref{eq2} and~\ref{eq3}. The Lundquist flux rope model implies increasingly varying $\tau$ from the axis, whereas the GH model gives a constant twist from centre to the boundary of the flux rope. This kind of $\tau$ variation was also found in some case studies of \citet{huqiang2014}. GH configuration is thus inferred to be more consistent with the general behaviour of field line turns in the GS result of the MC's magnetic structure.

Since the Lundquist fit infers the radius of MC as 0.135 AU, the twist number is (${{T}_{0}}=\frac{\alpha }{2\pi }=\frac{2.405}{0.135\times 2\pi }\text{turns/AU}$ )  estimated to be 2.83 turns/AU. In the GH fitting, the twist serves as an additional fitting parameter, so the fitting directly results in 1.2 turns/AU. Therefore, the result of about one turn from GH result is quite consistent for most of the GS twist (flat value of 1.75) results in Figure~\ref{Fig5}. This inference is based on one event, however a better fit, in most of the cases especially the rotating component, is generally expected from the GH model fitting than the Lundquist model fitting due to the internal field configuration. 

\subsection{Gold-Hoyle versus Lundquist models with expansion effect}
\begin{figure*}[!htp]
\centering
\includegraphics[width=.7\textwidth,clip=]{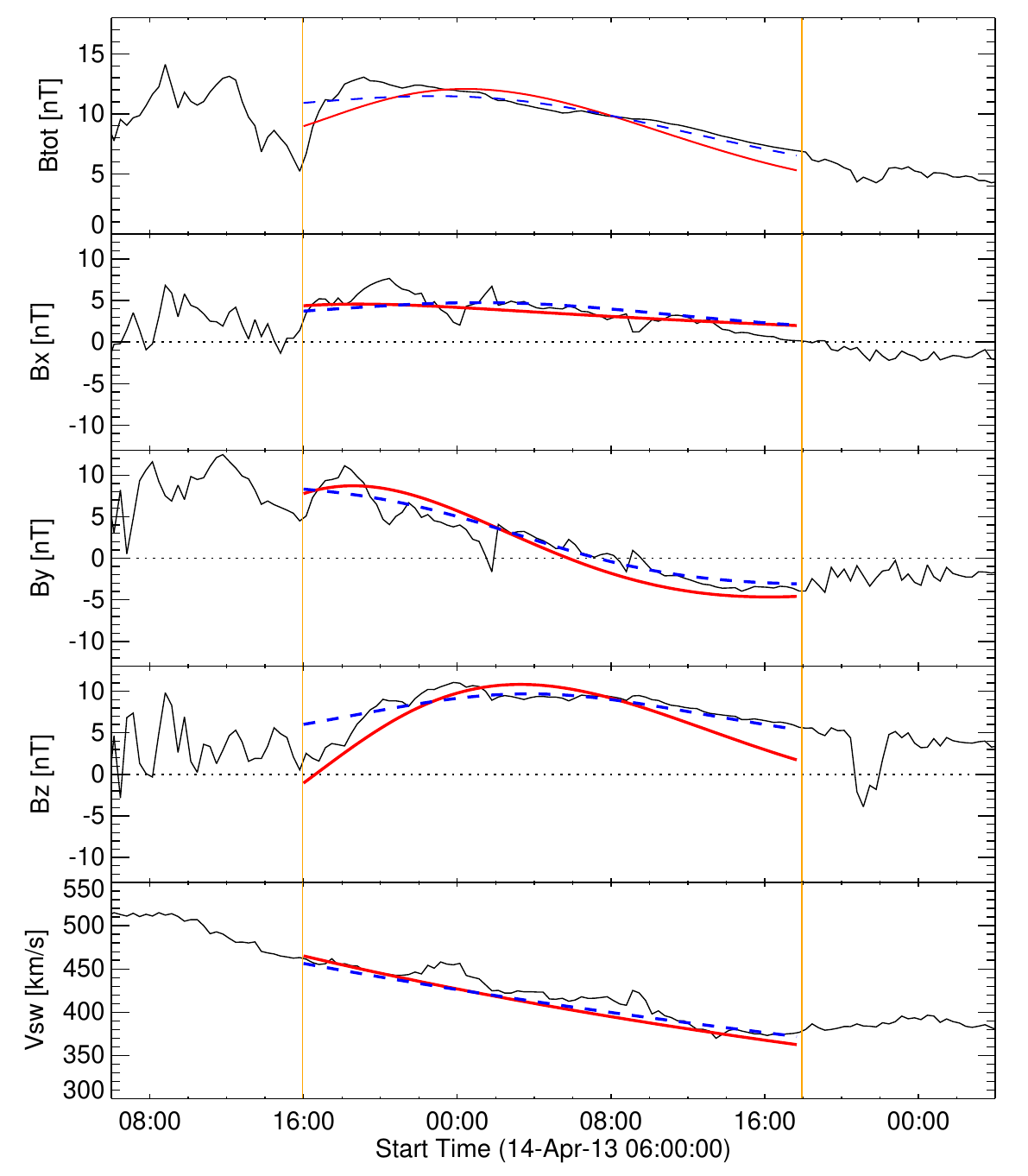}
\caption{Same as Figure~\ref{Fig6} but accounting the effect of self-similar expansion \citep{marubashi2007} in the MC fitting. $T_0=1.1$ turn/AU is used for the GH configuration.}
\label{Fig7}
\end{figure*}

While the spacecraft traverses the MC, it undergoes expansion significantly in a time scale of a day.  To account this expansion in the fitting, \citet{farrugia1992} proposed a self similar expansion model to the MC. This model was later applied to cylinder and torus geometry by \citet{marubashi2007}. In this model, the radius of the MC (cylinder, Equations~\ref{eq2} and~\ref{eq3}) varies with time (t) as $r(t)=r_0(1+Et)$ where E is the expansion rate, by which means force-free parameter  $\alpha$ is time varying while being spatially constant. And the magnetic field components also inversely proportional to $(1+Et)^2$. The model relies on the observed data of solar wind in addition to magnetic field components, and the MC fitting determines E, the average solar wind velocity $U_0$, along with other parameters as in the models without expansion. Following this fitting procedure described in \citet[Appendix A]{marubashi2007}, the results for GH and Lundquist configurations are plotted in Figure~\ref{Fig7}. And the fitted parameters are listed in Table~\ref{tab1}. 

During the fitting procedure, we set $U_0$ to the $V_{HT}$ frame velocity (414 km/s) for both the configurations. In the Lundquist configuration, the MC radius ($r_0=0.12$ AU) yields $T_0=3.19$ turns/AU. Whereas for GH configuration, the fitting converges (especially the rotating component) when $T_0=1.1$ turns/AU. This is a consequence of reducing twist number from the centre towards the MC boundary to a uniform level of 1.75 turns/AU and the expansion effect mimics it to reproduce the observations. The expansion coefficient (0.22 per day) is significant that the initial MC radius increases by 25\% at the time of the spacecraft passes the MC rear boundary. The orientation differs significantly in $\phi$ ($132^o$) with a similar $\theta$ as GS method. The rms deviation between the observed and model magnetic fields for the Lundquist and GH fits are 26.1, 1.91 respectively, which delineates the goodness of GH fit over the earlier. 

\section{MC fitting}
\label{sec5}   
\section{Comparison of magnetic properties in the MC and the Source AR }
\label{sec4}

To compare the AR twist with that of the MC, the total field line length is required, as direct comparison of twist parameter is not suitable because field lines stretch while flux rope structure expands. \citet{kahler2011a} derived total field line lengths by using solar energetic particles, like electrons, of known speeds. The procedure requires the observations of solar release times as bursts and their 1 AU onset times in the form of type III radio emission. Recently, \citet{huqiang2015} utilized these procedures and showed that the in situ measurements of field line lengths are consistent with a flux rope structure with spiral field lines of constant and low twist. Based on their analysis of a limited number of MC events, they argued that under most circumstances, the effective axial length of a cylindrical flux rope is $L_{eff}\in[1, 2]$ AU. They suggested to adopt this range of field line axial length for the relevant studies of deriving and relating various physical quantities to their solar sources.

\begin{table*}[!ht]
\centering
\begin{threeparttable}
\small
\centering
\caption{Comparison of GS Reconstruction results with the MC fitting results. The columns refer to the following parameters. $V_{HT}$: deHoffmannTeller frame velocity (km/s), $B_0$: field strength at the center of the cylinder (nT), $\theta$: latitude angle of MC axis (deg), $\phi$: longitude angle of MC axis (deg), $T_0$: twist number at the center (turns/AU), D: MC diameter (AU), p: impact parameter (AU), $\Phi_A$: axial flux ($10^{21}$ Mx), $\Phi_P$: poloidal flux ($10^{21}$ Mx), I: axial current (GA), $U_0$: velocity of solar wind (km/s), E: expansion rate (/day), Erms: error in root mean square between observed and modelled field.}
\begin{tabular}{l l l l l l l l l l l l l l}
\hline
Method	    &   $V_{HT}$	&  $B_0$   &  $\theta$   &	$\phi$	&  $T_0$	& D	& p	& $\Phi_A$	& $\Phi_P$	& I		& $U_0$ &	  E   & Erms \\
\hline
GS														& 414	& 	13.0		&	72			&	101		&	-9.0	&	0.26	& 	0.03	& 0.46	&	0.73	& 	0.477	&	--         & 	--	   &	--        \\
Lundquist									& 414	& 	13.0	&	73.8	&	73.7	&	-2.83	&	0.27	& 	0.04	& 0.71	&	1.63  & 	0.675	 &	-- & 	--	   &  0.266  \\
GH 														& 414	& 	12.2	&	74.6	&	74.4	&	-1.2	&	0.28	& 	0.05	  & 1.12	&	1.34	& 	0.637	&	--        & 	--	   &  0.176 	\\
Lundquist+Exp        & 414	&   17.3  &	69.8  &	132	  &	-3.19	&	0.24	& 	0.06	  & 0.76	&	1.94	& 	0.807  &	414 	 &  0.28 & 2.016    \\
GH+Expansion				& 414	&   14.1   &	72.6  &	62.5  &	-1.1	&	0.24	  & 	0.04	  & 1.1  &	1.19   & 0.621	 & 414      &  0.22 &	0.147	\\
\hline
		\end{tabular}
%\begin{tablenotes}
%\end{tablenotes}
\end{threeparttable}
\label{tab1}
\end{table*}

\begin{table*}[!ht]
\centering
\begin{threeparttable}
\caption{Comparison of magnetic twist in the MC and the Source AR}
\begin{tabular}{l l l}
\hline
AR twist at 06:50UT on 11 April 			&		$(\alpha L)_{AR}$ [L=103 Mm]					 	&  $(\alpha L)_{MC}$  [L=1-2 AU]  \\
\hline	
$\alpha_{av}=1.0\times 10^{-8}m^{-1}$\tnote{[a]}	    & 0.16 turns	    &   1.75-18.0 turns               \\
$\alpha_{av}=4.16\times10^{-8}m^{-1}$\tnote{[b]}	    & 0.7 turns	      &    1.75-18.0 turns      				\\
$\alpha ={{10}^{-7}}-{{10}^{-6}}{{m}^{-1}}$\tnote{[c]}	  & 1.6-16 turns  &    1.75-18.0 turns            	\\
\hline
		\end{tabular}
		\begin{tablenotes}
\footnotesize
\item[[a]] average over entire AR, i.e., field-of-view of Figure~\ref{Fig1}(a)
\item[[b]] average over sunspot region,i.e., inset in Figure~\ref{Fig1}(a)
\item[[c]] local distribution in the sunspot region
\end{tablenotes}
\end{threeparttable}
\label{tab2}
\end{table*}
Considering the field lines of total length $L=1-2$AU, the GS based MC twist distribution ($(\alpha L)_{MC}$) comes out to be 1.75-18.00 turns, reflecting the full variation from the center to the edge in Figure~\ref{Fig5}. Note that $(\alpha L)_{MC}$ can also be referred as $(\tau L)_{MC}$. Now in the source region, given the distance of 66 Mm (see Figure~\ref{Fig1}, and also, \citet{vemareddy2014b} between the legs of the sigmoidal flux rope, assuming a half torus shape, the average length of the field lines in the sigmoidal flux rope would be 103 Mm. The average twist from the sunspot, where one of the legs of sigmoidal flux rope lies, at the time of eruption is estimated as $-4.16\pm 0.32\times {{10}^{-8}}{{m}^{-1}}$. This results in an AR twist distribution ($(\alpha L)_{AR}$) of 0.7 turns which is less by a factor of three with the range of $(\alpha L)_{MC}$. Note a $2\pi$ factor when referring $(\alpha L)_{AR}$ in units of turns. We note that $\alpha$ values in the sunspot region are distributed in the range ($10^{-7}-10^{-6}$)m$^{-1}$. Since the twist distribution in the GS-based MC is derived from the local magnetic field distribution, we argue to use the local range of $\alpha$ in the source AR too. By doing so, we arrive $(\alpha L)_{AR}$ as 1.6-16 turns, which is well within the range of $(\alpha L)_{MC}$. Note that we followed the same procedures of \citet{leamon2004}, who found differing MC's twist distribution of an order compared to their source regions. They assumed $L=2.5$AU for the computation. Their result could likely be due to the use of average $\alpha$ over the entire AR and also less resolution magnetic field observations. Indeed, this is the case here (0.16 turns) for the entire AR value (${{\alpha }_{av}}=1.0\times {{10}^{-8}}{{\text{m}}^{-1}}$, see Table~\ref{tab2}). Note that high resolution, high sensitive magnetic field observations will always improve the AR twist estimation with higher magnitude. It is thus obvious that even the moderate values $0.5\times10^{-7} m^{-1}$ of AR twist distribution would be comparable with the most of the twist distribution in MC cross-section. 

The ratio of axial  flux ($0.46\times10^{21}$Mx) from the MC and from the source sunspot region ($3\times10^{21}$Mx) is 0.15. Similarly, the net current ratio is $0.34\times10^{-3}$. These values are typical and consistent with the cases presented in \citet{leamon2004}. From both the fittings with and without expansion, due to the increasing twist from center to the boundary, the Lundquist fit converges at higher value of twist number for the MC than that from GH fitting. As of this fact, the poloidal flux in Lundquist fitting is significantly higher (by a factor 2.5) than axial flux.

\section{Summary and Discussion}
\label{sec6}
We have analysed in situ observed MC and its solar source region in an effort to emphasize the flux rope connections of Sun and Earth. Magnetic flux ropes play prime role in the Sun-Earth connections during major solar eruptive events like CMEs. The MC structures are accepted to be the part of large scale bent flux rope with legs still having connections to the solar source AR. The solar AR 11719 is found to be the source region of the observed MC during April 14-15, 2013. The pre-eruptive AR has a well-developed inverse-S sigmoidal flux rope under the evolving conditions of cancelling and approaching flux regions. This sigmoidal flux rope erupted on 6:50UT on April 11, 2013 and launched a halo CME directed at an average speed of 861km/s towards Earth. 

Utilising the in situ magnetic field observations, we examined the magnetic structure of the flux rope by the GS reconstruction method.  The MC axis points at $72^o$ latitude ($\theta$ ) and $101^o$ longitude ($\phi$ ) in the GSE system, where the latitude determines the deviation of the MC axis from the zenith. Since the source region sigmoid is aligned at $225^o$ from the ecliptic north, a possible rotation of the apex of the flux rope (upto $117^o$, \citealt{green2007,lynch2009}) in its expansion during CME eruption, could result in such a predicted axis orientation of $72^o$ latitude. The axial field ($B_z$) in the MC structure is positive (northward) with a left handed twist consistent with the source region sigmoid morphology and magnetic field distribution.

This MC magnetic structure has a field line twist number of 9 turns/AU at the center, which is decreasing to a flat value of 1.75 turns/AU upto the boundary. As also found in a handful of cases by \citet{huqiang2014, huqiang2015}, this inferred field line turns from the GS method is more consistent with the constant twist GH configuration rather the Lundquist configuration of increasing twist from the axis of the flux rope to its boundary. Because of this reason, the GS magnetic structure is non-linear force-free and hence it would be more appropriate to use GH magnetic configuration to fit the in situ magnetic fields for estimating parameters of cylindrically symmetric MC structures. 

For more clues on the connections of source region, we compared source region magnetic properties with the GS based MC magnetic structure. The net absolute axial flux and vertical current from source sunspot region are comparatively small to that from MC. These values are typical and consistent with the cases presented in \citet{leamon2004}. In contrast to their findings, the magnetic twist of the pre-eruptive AR is comparably in the range of MC's twist number. Identifying the anchoring region of the flux rope foot point, high resolution and high sensitive magnetic field observations in the source region better quantify the twist that manifests the flux rope. On the other hand, the length of field lines in the MCs is better constrained [1-2 AU] now by previous studies (i.e. from \citealt{huqiang2015}) and we now have a better grasp of the MC structure. We suggest to consider the distribution of  $\alpha$ from the source region for its comparison with the in situ MC. 

The cylindrically symmetric Lundquist and Gold-Hoyle configurations reproduce the global trend of in situ magnetic field observations especially the rotating component. The resulting orientation of the flux rope for both the models is almost the same. The helicity sign H and the field strength $B_0$, inclination angle are in agreement with the GS reconstruction results. Due to the flat value of twist in the most part the MC, GH configuration with 1.1 turn/AU resembles the GS twist value and fits better the observations over the Lundquist configuration. In fact, these fitting results improved (in terms of Erms) when considered the self-similar expansion into account \citep{marubashi2007}. All the results including GS method yield the higher poloidal flux than the axial flux, indicating the twisted nature of field lines in the MC. Due to increasing twist profile from center to boundary, the Lundquist fit estimates higher twist and poloidal flux than the GH fit. The GS reconstruction provides clues on the MC twist structure, which for the first time, fitted with GH configuration as an alternative to general practice of Lundquist configuration. Although the fitting improved to a great extent, it is yet to see, in a large sample of data sets, the goodness of GH fitting (including the expansion effect) in characterising the properties of in situ MC structures and its source region links. %Because the pre-formed flux rope structures support the plasma material in the dips of wounded field lines, filament eruptions, as is the case here, are suggested to associate with MCs of higher twist (more than 3 turns/AU) than those from prominence eruptions \citep{huqiang2014}.

In conclusion, the connection of the active region to the magnetic cloud is unambiguous. The length of the field lines, both in the MC and in the source region sigmoid, is better constrained for a quantitative comparison of source region magnetic signatures in the in situ MC. The fitting procedures determines the lower twist values away from the MFR center. These are the important points that have hindered in the previous studies on definite conclusions of how the twist behaves in magnetic flux ropes on the Sun and at 1 AU. For such twisting behaviour, this study suggests an alternative fitting procedure to better characterise the MC magnetic structure. These two points opens up the possibility to predict the MC twist configuration from the solar imaging information, which is of very high relevance to understanding the origin of the CME magnetic field (and its $B_z$) and thus space weather prediction in general. 

\acknowledgements SDO is a mission of NASA's Living With a Star Program. We thank the referee for the suggestions which helped to improve the paper. P.V. is supported by an INSPIRE grant of AORC scheme under the Department of Science and Technology.  W.M is funded by Chinese Academy of Sciences President's International Fellowship Initiative Grant No. 2015PE015, NSFC Grants No. 41131065 and 41574165. C.M. and T.R. thank the Austrian Science Fund (FWF): [P26174-N27]. The presented work has received funding from the European Union Seventh Framework Programme (FP7/2007-2013) under grant agreement No. 606692 [HELCATS]. CJF was partially funded by STEREO FARSIDE grant to UNH and NASA STEREO grant NNX15AU01G (decline of cycle 24). We thank the ACE SWEPAM and MAG instrument teams and the ACE Science Center for providing the ACE data. 

\bibliographystyle{apj}
%\bibliography{ref}

%%%%%%%%%%%%%%%%%%%%%%%%

\end{document}